\documentstyle[preprint,aps,prd]{revtex}
\tightenlines
\newcommand{\be}{\begin{equation}}
\newcommand{\br}{\begin{array}}
\newcommand{\er}{\end{array}}
\newcommand{\beq}{\begin{equation}}
\newcommand{\ee}{\end{equation}}
\newcommand{\eeq}{\end{equation}}

\newcommand{\<}{\langle}
\newcommand{\D}{\Delta}
\newcommand{\G}{\Gamma}
\newcommand{\p}{\partial}

\renewcommand{\>}{\rangle}
\def\ba{\begin{eqnarray}}
\def\ea{\end{eqnarray}}

\begin{document}

\preprint{hep-th/9911073}
\title {The propagator for a general form field in $AdS_{d+1}$}
\author{Iosif Bena}
\address{University of California, Santa Barbara, CA 93106 \\ iosif@physics.ucsb.edu}
\date{\today}
\maketitle

\begin{abstract}
Using the known propagator equations for 0,1 and 2 forms in $AdS_{d+1}$, we find the $p$-form field propagator equations in dimensions where the forms are Poincar\'e dual. Since the general equation obeyed by the propagators is linear in dimension, this gives us the equation obeyed by the propagators for any $d$. 
Furthermore,  based on the Poincar\'e duality formulas for 0,1,2 and 3 forms we conjecture the general form of the Poincar\'e duality formulas, and check them against the previously found propagator equations. The whole structure is self-consistent. Once we have the equations, we can easily obtain all the $p$-form field propagators in $AdS_{d+1}$. The generalization to massive $p$-forms can be also easily done.
\end{abstract}
\pacs {11.25.-w; 04.50.+h }

\section{Introduction}

In \cite{fred,df1,df2,io} the propagators for scalars, vectors and antisymmetric 2-tensors in $AdS_{d+1}$ were computed. Furthermore, in \cite{io} Poincar\'e duality between the vector and 2-tensor propagator was verified for $d=4$. 
In a maximally symmetric space like $AdS_{d+1}$ the general propagator for a $p$-form field is a maximally symmetric bitensor.  As explained in \cite{allen}, this bitensor is a product of antisymmetrized partial derivatives of the chordal distance $u$ multiplied by a scalar function $G_p(u)$. The form of the equations obeyed by $G_p$ is very similar for the cases we know ($p$=0,1, and 2).  By using Poincar\'e duality between the 3-form field and the 0,1 and 2 - form fields for $d$=4,5 and 6 we can find the form of the equations obeyed by $G_3$ in these dimensions. Since the equations are obtained from the equations of motion, and by contracting at most 2 derivatives acting on chordal distances, they must depend on the dimension of the space in a linear way. Thus, it becomes very easy to see the general form of the equation of motion obeyed by $G_3$. The same method can be easily generalized for all the $p$-form propagators.

\section{ The 2-form propagator - brief review}

In this section we will be giving an outline of the method used in \cite{io}. In Euclidean $AdS_{d+1}$, with the metric
$$ds^2={1 \over z_0^2}(dz_0^2+\Sigma_{i=1}^d{dz_i^2} ), \eqno(1)$$
the easiest way to express invariant functions and tensors is in terms of the chordal distance:
$$u \equiv {(z_0-w_0)^2+(z_i-w_i)^2 \over 2 z_0 w_0}. \eqno(2)$$
We work in the subspace of conserved currents, so gauge fixing is not necessary. 

The 2-form propagator, $G_{\mu\nu;\mu'\nu'}$ is a bitensor, and as explained in \cite{allen,fred,io}, it can be expressed as a combination of 2 bitensors:
$$ T^1_{\mu\nu;\mu'\nu'} = \p_{\mu}\p_{\mu'}u  \p_{\nu} \p_{\nu'}u -\p_{\mu} \p_{\nu'}u \p_{\nu}\p_{\mu'}u \eqno (3)$$
$$ T^2_{\mu\nu;\mu'\nu'} = \p_{\mu}\p_{\mu'}u  \p_{\nu}u \p_{\nu'}u -\p_{\mu}\p_{\nu'}u \p_{\nu}u \p_{\mu'}u - \p_{\nu}\p_{\mu'}u \p_{\mu}u \p_{\nu'}u +\p_{\nu}\p_{\nu'}u \p_{\mu}u \p_{\mu'}u $$
Nevertheless, it is more advantageous to decompose it as
$$G_ {\mu\nu;\mu'\nu'} = T^1_{\mu\nu;\mu'\nu'} G_2(u) + D_{\mu}V_{\nu;\mu'\nu'}- D_{\nu}V_{\mu;\mu'\nu'}, \eqno(4) $$
where the second part is a gauge artifact.
Upon using the equations of motion, which involve 2 covariant derivatives applied on a combinations of  $G_{\mu \nu; \mu' \nu'}$'s, we find the equation of motion obeyed by $G_2$ by equating the scalar coefficients multiplying  $T^1$ and $T^2$. In doing this we use the formulas in the appendix. The contractions of two covariant derivatives which these formulas allow us to do, give us extra factors multiplying G and its derivatives. Nevertheless, there is only one contraction per term, and thus these extra factors in the equation for $G_2$ can only be linear in $d$. We obtain an equation of a hypergeometric form, whose solution can be easily found.

\section{The tensor structures of the p-form propagators}
Similarly to the 2-form, the 3 form propagator $G_{\mu\nu\rho;\mu'\nu'\rho'}$ is antisymmetric in both primed and unprimed indices, and can be expressed as a sum of 2 bitensors similar to $T^1$ and $T^2$ multiplied by functions of $u$. Nevertheless, it is more convenient to combine the 2 bitensors into a gauge invariant part and a gauge artifact:
$$G_{\mu\nu\rho;\mu'\nu'\rho'} = (-1)^3[\p_{\mu}\p_{\mu'}u \p_{\nu}\p_{\nu'} u  \p_{\rho}\p_{\rho'}u + \mu'\nu'\rho' \ \hbox{antipermutations}] G_3(u) +D_{[\mu}   J_{\nu\rho]; \mu'\nu'\rho'},    \eqno(5)$$ 
where the square brackets denote antisymmetrization, the $(-1)^3$ has been put in to outset the $-1$ from (A1), and  $J_{\nu\rho; \mu'\nu'\rho'}$ is a bitensor antisymmetric in both primed and unprimed indices . Generalizing to higher form propagators is obvious. Their gauge invariant part is of the form: 
$$ G_{\mu_1 ... \mu_p; \mu'_1 ... \mu'_p }^{\hbox{\tiny{invariant}}}  = (-1)^p[\p_{\mu_1}\p_{\mu'_1}u ... \p_{\mu_p}\p_{\mu'_p}u + \hbox{antipermutations of primed indices}] G_p(u). \eqno(6)$$
The problem reduces therefore to finding the functions $G_p$  appearing in the propagators.
From \cite{fred,df1,df2,io} we know that the equations satisfied by $G_0,G_1$ and $G_2$ for $u \neq 0$ are:
$$ u(2+u)G_0''+(d+1) (1+u)G_0'+0(d-0)G_0=0, \eqno( 7a)$$
$$ u(2+u)G_1''+(d+1) (1+u)G_1'+1(d-1)G_1=0, \eqno( 7b)$$
$$ u(2+u)G_2''+(d+1) (1+u)G_2'+2(d-2)G_2=0, \eqno( 7c)$$
where we arranged the last term knowing in advance how the equation for $G_p$ will look like. There is another $\delta$ function term in the right hand side, which insures that we use the solution of (7a,b,c) which diverges as $u \rightarrow 0$, and gives the appropriate normalization to the solution. We will be discussing this in Chapter VI.
We can now use Poincar\'e duality to obtain the relations between $G_p$ and $G_0, G_1$ and $G_2$.

\section{Poincar\'e duality}
In $d+1$ dimensions with Euclidean signature, a $p$ form field is Poincar\'e dual to a $d-p-1$ form field by the relation
$$F_{\mu_1...\mu_{p+1}} \epsilon^{\mu_1\mu_2 ... \mu_{d+1} }=i (p+1)! \ F^{\mu_{p+2}...\mu_{d+1}},  \eqno(8)$$
where the $F$'s are the field strengths. Thus 
$$- \<{F_{\mu_1...\mu_{p+1}}(z) F_{\mu'_1...\mu'_{p+1}}(w) }\> \epsilon^{\mu_1\mu_2 ... \mu_{d+1} } \epsilon^{\mu'_1\mu'_2 ... \mu'_{d+1} } = ((p+1)!)^2 \<{ F^{\mu_{p+2}...\mu_{d+1}}(z) F^{\mu'_{p+2}...\mu'_{d+1}}(w)}\>. \eqno(9)$$
Technically speaking, (9) contains also a contact term, but we will be concentrating for now on $u \neq 0$ (chapters IV and V), and take it into account later (chapter VI). The correlation function of the field strengths is obtained by appropriately differentiating the propagator.

If we apply Poincar\'e duality naively, we expect to obtain a relation between $G_p$ and $G_{d-p-1}$. Indeed, in flat space, the propagators are dual. On the other hand,  $AdS_{d+1}$ behaves like a ``box''; more precisely, it is necessary to specify boundary conditions at $\infty$ for the solutions of the propagator equations. Since the  boundary conditions will not be in general invariant under the duality, the propagators will not be dual in general (we will comment on the duality between boundary conditions at the end of Chapter VI). Nevertheless, a particular combination of the 2 solutions of one equation and of the 2 solutions of the other will be. Which combination it is depends on the particular duality we consider. In the case of a duality between a $p$ form and a $q$ form, we will use $H_p$ to denote the solution of the equation for $G_p$ which satisfies this duality. We are not interested in what particular combination of he solutions $H_p$ is; the only thing which we need to know about $H_p$ is that it satisfies the same equation as $G_p$.

In the case of the 0-form, we compare $\<{F_{z_0} F_{z_0'} }\>$ with its left hand side counterpart in (9). This insures that only indices different from $z_0$ and $z_0'$ are summed over, which simplifies the computation very much. For $d=p+1$ we obtain from (6) and (9) after a few steps:

$$(2u+2)H_p''+d H_p'-H_p''({w_0\over z_0}+ {z_0\over w_0})=H_0''(u^2+2u+2)+H_0'(u+1)-( {w_0\over z_0}+ {z_0\over w_0})((1+u)H_0''+H_0'). \eqno(10)$$
Since $ ({w_0\over z_0}+ {z_0\over w_0}) $ can change with $u$ constant and vice versa, (10) implies: 
$$(2u+2)H_p''+d H_p'= H_0''(u^2+2u+2)+H_0'(u+1) \eqno (11)$$
$$H_0''(1+u)+H_0' = H_p'' \ \ \ \hbox{for} \ \ d=p+1. \eqno(12) $$
Combining (12),(11), and the equation for $G_0$ (7a) which $H_0$ also satisfies, we obtain:
$$H_0'(1+u) = H_p' \ \ \ \hbox{for} \ \ d=p+1. \eqno(13a) $$
In the case of the 1 form and 2 form we compare $\<{F_{z_0z_i} F_{z_0'z_i'} }\>$, respectively $\<{F_{z_0z_iz_j} F_{z_0'z_i'z_j'} }\>$ with their counterparts in (9). We obtain after similar steps
$$H_1'(1+u)+H_1 = H_p'  \ \ \ \hbox{for} \ \ d=p+2, \eqno(13b) $$
$$H_2'(1+u) + 2 H_2 = H_p'  \ \ \ \hbox{for}\ \  d=p+3. \eqno(13c) $$
Having these equations enables us to find the propagator equations.

\section{The p-form propagator equations}

We first examine the 3-form propagator. Using  (13a,b,c) and combining them with (7a,b,c) we find the equations $G_3$ satisfies for $d$=4,5 and 6, $u \neq 0$ to be
$$ u(2+u)G_3''+5(1+u)G_3'+3G_3=0 \ \ \ \hbox{for} \ \  d=4  \eqno(14a)  $$
$$ u(2+u)G_3''+6(1+u)G_3'+6G_3=0 \ \ \ \hbox{for} \ \  d=5 \eqno(14b) $$
$$ u(2+u)G_3''+7(1+u)G_3'+9G_3=0 \ \ \ \hbox{for} \ \  d=6. \eqno(14c) $$
To be more precise, (14a,b,c) are actually the equations satisfied by $H_3$ in these dimensions. One legitimate objection one may have is - are these equations the ones satisfied by $G_3$, or are they just some other equations satisfied by the particular combinations of solutions which respect Poincar\'e  duality. In Appendix II we will address this objection and show that these equations are indeed the ones satisfied by $G_3$.

Based on the arguments presented above that the dimension only enters the equation for $G_p$ linearly, the equation $G_3$ satisfies is:
$$ u(2+u)G_3''+(d+1)G_3'+3(d-3)G_3=0.  \eqno(15) $$

We can use the same procedure for the $p$-form in $d=p+1,d=p+2$ and $d=p+3$, to find the equation for $G_p$ to be:
$$ u(2+u)G_p'' + (d+1)(1+u)G_p' +p(d-p)G_p=0 . \eqno(16) $$
This reduces to (7a,b,c) for the 0,1 and 2 forms.
Once we have this equation it is very easy to find the propagators. 

We can check whether the propagator equations we have are correct by using Poincar\'e duality between the 3 form field and a higher $d-4$ form field in $AdS_{d+1}$. The equations for both these propagators was found using our slick argument, and not by cranking through the large number of terms in the equation of motion.
By comparing  $\<{F_{z_0z_iz_jz_k} F_{z_0'z_i'z_j'z_k'} }\>$, with its counterpart in (9) we obtain:
$$(1+u)H_3'+3H_3=H_{p}'  \ \ \ \hbox{for}\ \  d=p+4. \eqno(17)$$
It is very easy to check that using (17) and equation (15) for $G_3$ we obtain the equation (16) for $G_{d-4}$. Thus, our approach is consistent.
Now, if we look at the relations derived from Poincar\'e duality between $H_p$ and $H_0, H_1, H_2 $ and $H_3$ (11a,b,c,16), it is impossible not to observe a pattern. Namely 
$$(1+u)H_q'+q H_q=H_p'\ \ \ \hbox{for} \ \ d=p+q+1. \eqno(18)$$

We can check this guess by plugging it in the equation for $G_p$ and obtaining the equation for $G_q$, and vice versa.
As we can see, we uncovered the whole structure of relations between the propagator equations and the Poincar\'e duality formulas.
 
\section{The propagators}

Equation (16) is a second order hypergeometric differential equation, which has 2 solutions. Near $u=0$ the solutions go as $f_1=c,\ \ \  f_2={1\over u^{d-1\over 2}}$. As $u \rightarrow \infty$ the solutions go as  $f_3={1\over u^p },\ \ \  f_4={1\over u^{d-p}}$ for $d\neq2p$, and as  $f_3={1\over u^p },\ \ \  f_4={\log u \over u^{p}}$.
As we discussed, in Chapter III, there is another $\delta$ function term in (7a,b,c), which comes from the equations for $ G_{\mu_1 ... \mu_p; \mu'_1 ... \mu'_p } $ (equation (6) in \cite{io} for the 2 form, equation (2.5) in \cite{fred} for the 1 form). We can see in the case of a general form that the  $\delta$ function appears in the equation for  $ G_{\mu_1 ... \mu_p; \mu'_1 ... \mu'_p } $ multiplying an antisymmetrized product of metrics, and that its sign will be opposite to the sign of $u(u+2) G_p''$. 

We expect the physical solution to be the one which diverges at the origin because of the $\delta$ function source.  Moreover, since we are in $AdS$ we impose the boundary condition that the solution decays the fastest as $u \rightarrow \infty$. 

Using some hypergeometric ``functionology'', we can see that the solution which decays the fastest at $\infty$ diverges at the origin, and that the one which is constant at the origin decays the slowest at  $\infty$.
Thus, the solution which decays the slowest as $u \rightarrow \infty$ is excluded.

There are 2 different cases, based on the relation between $d$ and $p$.
For $d-1<2p $ the solution is a hypergeometric function:
$$G_p =  {\G({(d-1)/2})\over 4 \pi^{(d+1)/2} u^{d-1\over 2} } {c \ F({2p+1-d \over 2},1+p-d,2p+1-d,{2\over 2+u})\over (2+u)^{2p+1-d \over 2}}, \eqno(19)$$
where $c$ is a constant that normalizes the second fraction to 1 as $u \rightarrow 0$.
The first fraction is a normalization constant which takes care of the $\delta$ function discussed above. The hypergeometric function has the third coefficient negative, and can be simplified to a rational function.

For $d+1>2p$ the solution is also hypergeometric:
$$G_p =  {\G({(d-1)/2})\over 4 \pi^{(d+1)/2} u^{d-1\over 2} }
{c \ F({d+1-2p \over 2},1-p,d+1-2p,{2\over 2+u})\over (2+u)^{d+1-2p \over 2}}, \eqno(20)$$
where c is a constant similar to the previous case. This solution can also be simplified to a rational function. The ranges of validity overlap at $d=2p$, where the solutions also overlap.

As promised, we now come back to the invariance of boundary conditions under Poincar\'e duality. Equation (18) is obeyed by our solutions for $p>q$ only (it is easy to see this by looking at their behavior at $\infty$). Equation (18) was obtained by contracting with the $\epsilon$ tensor the indices of the p-form propagator which were different from $z_0$. We therefore conclude that the boundary conditions for the components of the higher form field strength which do not contain $z_0$ are dual to the boundary conditions for the components of the lower form field strength which contain $z_0$.

\section{The massive case}
It is very easy to generalize our results (19,20) to the case of a massive form field. Adding a mass term to the $p$-form action spoils its gauge invariance. Thus, we cannot add in the equation for the propagator (B1) terms which cancel when integrated against a covariantly conserved current. 

The tensorial structure (4,5) of the propagator is not changed by adding a mass term. What changes is that the second part of the propagator ( $ D_{[\mu_1} V_{\mu_1 ... \mu_{p-1}]; \mu'_1 ... \mu'_{p}} $ - the antisymmetrized covariant derivative acting on a $(p,p-1)$ antisymmetric bitensor) can no longer be interpreted as a gauge artifact. It is a legitimate term in the propagator which we have to find.

A mass term in the Lagrangian introduces another term in the left hand side of the equation for the propagator (B1): $m^2 G_{\mu_1 ... \mu_p; \mu'_1 ... \mu'_p } $. The second part of the propagator cancels in the term with 2 covariant derivatives acting on the propagator. Multiplied by $m^2$ it also takes the place of $D_{[\mu_1'} \Lambda_{\mu_1 ... \mu_p; \mu'_1 ... \mu'_{p-1}]}$ in (B1).

Thus, the only change in the equation for $G_p$ will be another mass term. Equation (16) will become:
$$ u(2+u)G_p'' + (d+1)(1+u)G_p' +[p(d-p)-m^2] G_p = 0 . \eqno(21) $$
Again the solution which decays the fastest at $\infty$ diverges at the origin, and the solution which is finite at the origin decays the slowest at  $\infty$. Thus, the solution which decays the slowest as $u \rightarrow \infty$ is excluded. The solution is a hypergeometric function:

$$G_p= {\G({(d-1)/2})\over 4 \pi^{(d+1)/2} u^{d-1\over 2} }
{c \ F(\D+{1 \over 2},\D+1-{d \over 2},2\D+1,{2\over 2+u})\over (2+u)^{d+1-2p \over 2}}, \eqno(22) $$
where $\D \equiv \sqrt{({d\over 2} - p)^2+m^2}$, and c is a constant similar to the one in (19) and (20). Equation (22) reduces to (19) and (20) for $m^2=0$.

What is now left is finding $ D_{[\mu_1} V_{\mu_2 ... \mu_{p}]; \mu'_1 ... \mu'_{p}} $. Since $V_{\mu_2 ... \mu_{p}; \mu'_1 ... \mu'_{p}}  $ is a bitensor, it can be expressed as
$$V_{\mu_2 ... \mu_{p}; \mu'_1 ... \mu'_{p}} = F_p(u)[\p_{[\mu'_1}u \ \p_{\mu_2}\p_{\mu'_{2}}u \  ... \ \p_{\mu_{p}}\p_{\mu'_{p}]}u ], \eqno(23) $$
where the square brackets denote antisymmetrization of primed indices.
By the arguments in Appendix B we know the relation between $F_p$ and $G_p$ to be of the form: 
$$(u+1)G_p'+A G_p= B F_p. \eqno(24) $$
Nevertheless, we do not know the constants $A$ and $B$. One way to find them is to find all the terms coming from the equation for the propagator. This is what we have tried to avoid so far. Hopefully, there is another way which is easier. The propagator for a massive form satisfies another equation:
$$D^{\mu_1} G_{\mu_1 ... \mu_p; \mu'_1 ... \mu'_p }=0.  \eqno(25) $$
Using (6) and (23) we obtain after a few steps:
$$u(2+u) F_p''+(d+3)(1+u) F_p' + p(d-p+2)F_p+(-1)^p[(1+u)G_p'+(d-p+2)G_p]=0. \eqno(26)$$ 
Combining (24) with (26) and (21) we obtain:
$$(u+1)G_p'+(d-p)G_p= {(-1)^{p+1} \over m^2} \ F_p. \eqno(27)$$

It is interesting to observe how much was obtained using just Poincar\'e duality and a few general properties of propagators in $AdS$.

While we were working on Chapter VII (which contains the generalization of our results to massive forms, and which did not appear in the first version of this preprint), \cite{asad} appeared, in which the massless and the massive form propagators were found using the equation of motion for the $p$-forms. Our results (21,27) were obtained in a different way, and agree with \cite{asad}.

 {\bf Acknowledgements:} I'd like to acknowledge useful conversations with Joe Polchinski, Costin Popescu, Gary Horowitz and Aleksey Nudelman. This work was supported in part by NSF grant PHY97-22022.

\appendix
\section{Several useful identities involving the chordal distance}

In the computations the following identities were useful:

$$ \p_{\mu} \p_{\nu'}u = {-1 \over z_0 w_0}[\delta_{\mu\nu'}+{(z-w)_\mu \delta_{\nu'0}\over w_0}+	{(w-z)_\nu' \delta_{\mu0}\over z_0} - u\delta_{\mu 0}\delta_{\nu' 0}] \eqno(A1)$$
$$ \p_{\mu} u = {1 \over z_0}[(z-w)_{\mu}/w_0-u \delta_{\mu 0}]	\eqno(A2)$$
$$ \p_{\nu'} u = {1 \over w_0}[(w-z)_{\nu'}/z_0-u \delta_{\nu' 0}] \eqno(A3)$$
$$ D^{\mu}\p_{\mu}u=(d+1)(u+1)					 \eqno(A4)$$
$$ 	\p^{\mu}u\  \p_{\mu}u = u(u+2)				\eqno(A5)$$
 $$ 	 D_{\mu}\p_{\nu}u	=g_{\mu\nu}(u+1)		\eqno(A6)$$
$$ 	(\p^{\mu}u)( D_{\mu}\p_{\nu}\p_{\nu'} u) = \p_{\nu}u \p_{\nu'} u 		\eqno(A7)$$
$$ 	(\p^{\mu}u)( \p_{\mu}\p_{\nu'}) u =( u+1) \p_{\nu'} u			\eqno(A8)$$
$$ 	 D_{\mu}\p_{\nu}\p_{\nu'} u = 		g_{\mu\nu}  \p_{\nu'} u			\eqno(A9)$$
 
\section{On the link between the equation for $H_p$ and the one for $G_p$}

We know that $H_p$ is a combination of the 2 solutions of the equation for $G_p$. Nevertheless, if we know the equation satisfied by $H_p$ it is not in clear that this is the equation $G_p$ satisfies.
We will try to determine under what conditions can $H_p$ satisfy 2 different equations of a similar form. In order to do this we need to go a bit through the gory details associated with finding the equation for $G_p$.

We will review the general procedure of obtaining the equation for $G_p$. The reader who is interested in more details can look up the derivations for $p=1$ and $p=2$ in \cite{fred} respectively \cite{io}.

The equation $G_p$ satisfies comes from the equation for $ G_{\mu_1 ... \mu_p; \mu'_1 ... \mu'_p }$, which for $u\neq 0$ is :
$$D^{\lambda} D_{[\lambda}  G_{\mu_1 ... \mu_p]; \mu'_1 ... \mu'_p } - D_{[\mu_1'} \Lambda_{\mu_1 ... \mu_p; \mu'_1 ... \mu'_{p-1}]}=0, \eqno(B1) $$
where the square brackets denote antisymmetrization and $\Lambda_{\mu_1 ... \mu_p; \mu'_1 ... \mu'_{p-1}}$ is a diffeomorphism whose contribution vanishes when integrated against a covariantly conserved current. The gauge non-invariant part of the propagator does not appear in the equation because it is itself the exterior derivative of a current.

Similar to the case of 2-form propagators, there are 2 bitensors which form a basis for antisymmetric bitensors:
$$T^1_{\mu_1 ... \mu_p; \mu'_1 ... \mu'_p } = \p_{\mu_1}\p_{\mu'_{1}}u ...  \p_{\mu_p}\p_{\mu'_{p}}u + \hbox{antipermutations of primed indices} \eqno(B2)$$

$$T^2_{\mu_1 ... \mu_p; \mu'_1 ... \mu'_p } = \p_{\mu_1}u \ \p_{\mu'_{1}}u \ \p_{\mu_2}\p_{\mu'_{2}}u ...  \p_{\mu_p}\p_{\mu'_p}u + \hbox{antipermutations of all indices}.$$  
Also, $\Lambda_{\mu_1 ... \mu_p; \mu'_1 ... \mu'_{p-1}}$ is a bitensor and can be expressed as:
$$\Lambda_{\mu_1 ... \mu_p; \mu'_1 ... \mu'_{p-1} } =A(u)[\p_{\mu_1}\p_{\mu'_{1}}u ... \p_{\mu_{p-1}}\p_{\mu'_{p-1}}u \ \p_{\mu_p}u + \hbox{antipermutations of unprimed  indices}], \eqno(B3)$$
where $A(u)$ is a scalar function.

It is possible to express (B1) as a sum of 2 expressions multiplying these 2 bitensors being equal to 0. These 2 expressions must be individually equal to 0. Since they contain $G_p(u)$ and $A(u)$, we can eliminate $A(u)$ and obtain an equation for $G_p$.

Using (6), we can see that the expression multiplying $T^1$ will contain:

- $u(u+2)G_p''$ coming from $\p_{\lambda}u\ \p^{\lambda}u\  G_p'' $ (using A5);

- $(u+1)G_p'$ coming from $\p_{\mu_i} \p^{\lambda}u \ G_p'$ and $\p_{\lambda} \p^{\lambda}u \ G_p'$ (using A4);

- $G_p$  coming from the 2 covariant derivatives acting on factors of the form $\p_{\mu_i}\p_{\mu'_{j}}u$ (using A9);

- $A(u)$ coming from the partial derivative acting on  $\p_{\mu_p}u$.

Thus, the general equation obtained from $T^1$ will look like:
$$u(u+2)G_p''+ B(u+1) G_p' + C G_p = D A(u), \eqno  (B4)$$
where $B,C,D$ are constants.
Similarly, the expression multiplying $T^2$ will contain:

- $(u+1)G_p''$ coming from $\p_{\lambda} \p_{\mu'_j}u \ \p^{\lambda}u \ \p_{\mu_i}u \ G_p'' $ (using A8);

- $G_p'$ coming from $D_{\lambda} \p_{\mu_i}\p_{\mu'_{j}}u \ \p^{\lambda}u \ G_p'$ (using A7);

- $A'(u)$ coming from the partial derivative acting on $A(u)$.

Thus the general equation  obtained from $T^2$ will look like:
$$(u+1)G_p''+ K G_p'  = L A'(u), \eqno  (B5)$$
with $K$ and $L$ constants. We can integrate (B5) once, set the integration constant to 0 (so that both sides vanish as $u \rightarrow \infty$), and combine it with (B4) to obtain the general form of the equation for $G_p$:
$$u(u+2)G_p''+ M(u+1) G_p' + N G_p = 0, \eqno  (B6)$$
where $M$ and $N$ some constants. If we had gone through the large number of terms in (B1) we would have determined $M$ and $N$.

We are now ready to explore the logical possibility that (14a,b,c) are just equations satisfied by the particular combinations of solutions giving $H_p$ and not the equations satisfied by $G_p$.

If $G_p$ satisfies another equation, it must be of the form (B6) and $H_p$ must also satisfy it. Since $H_p$ satisfies 2 different (B6)-like equations, it must be of the form $H_p = (1+u)^{\alpha}$, where $\alpha$ is any number. Clearly $(1+u)^{\alpha}$ cannot satisfy (14a,b,c),(15) or (16) for $p>2$ (which is the case of interest). Therefore (14a,b,c) are indeed the equations satisfied by $G_3$ Q.E.D.

\end {document}